\documentclass[conference]{IEEEtran}
\usepackage{amsmath,amssymb,amsfonts}
\usepackage{algorithmic}
\usepackage{graphicx}
\usepackage{textcomp}
\usepackage{xcolor}
\usepackage[utf8]{inputenc}
\usepackage[english]{babel}
\def\BibTeX{{\rm B\kern-.05em{\sc i\kern-.025em b}\kern-.08em
    T\kern-.1667em\lower.7ex\hbox{E}\kern-.125emX}}

\usepackage[
backend=biber,
style=numeric,
maxbibnames=99,
]{biblatex}
\usepackage[ruled,vlined]{algorithm2e}
\usepackage{hyperref} 
\usepackage{textcomp}

\begin{document}

\title{\textbf{COV19IR : COVID-19 Domain Literature Information Retrieval}}

\author{
    \IEEEauthorblockN{\textbf{\Large{Arusarka Bose}}\IEEEauthorrefmark{1},
    \textbf{\Large{Zili Zhou}}\IEEEauthorrefmark{2},
    \textbf{\Large{Guandong Xu}}\IEEEauthorrefmark{3}}
    \IEEEauthorblockA{\IEEEauthorrefmark{1}Indian Institute of Technology Kharagpur, India\\
    arusarka.bose@iitkgp.ac.in}
    \IEEEauthorblockA{\IEEEauthorrefmark{2}Department of Computer Science, University of Manchester, UK\\
    zili.zhou@manchester.ac.uk}
    \IEEEauthorblockA{\IEEEauthorrefmark{3}Advanced Analytics Institute, University of Technology Sydney, Australia\\
    Guandong.Xu@uts.edu.au}
}

\maketitle

\begin{abstract}
Increasing number of COVID-19 research literatures cause new challenges in effective literature screening and COVID-19 domain knowledge aware Information Retrieval.
To tackle the challenges, we demonstrate two tasks along with solutions, COVID-19 literature retrieval, and question answering.
COVID-19 literature retrieval task screens matching COVID-19 literature documents for textual user query,
and COVID-19 question answering task predicts proper text fragments from text corpus as the answer of specific COVID-19 related questions.
Based on transformer neural network, we provided solutions to implement the tasks on CORD-19 dataset, we display some examples to show the effectiveness of our proposed solutions.
\end{abstract}


\section{\textbf{\Large{Introduction}}}

Recently, increasing number of research works targeting COVID-19 contribute rich literature and base for further research in the area.
However, large data volume brings out new challenges, 1) effectively screening literature according to the user search query, 2) considering medical and COVID-19 domain knowledge in the information retrieval process. In this paper, we demonstrate our work on COVID-19 domain-specific information retrieval (IR) study. Two specific tasks are presented in this work, 1) COVID-19 literature retrieval, and 2) COVID-19 question answering. Our solutions on these two tasks are also given.

Textual retrieval task is a common task in IR area, searching the matching documents for each textual query provided by user. The COVID-19 literature retrieval task in this paper is a special textual retrieval task. It uses COVID-19 literatures as candidate documents set, the COVID-19 domain knowledge acts as important inter-document connections.
Besides that, the task orients to search queries given by professional researchers which may contain medical or COVID-19 related terminologies.
Our target is to link documents and queries based on COVID-19 domain knowledge aware semantic matching model.

COVID-19 question answering task is a new attempt of Natural Language Understanding (NLU) on COVID-19 literature dataset.
Treating COVID-19 literature dataset as an unstructured textual knowledge vault, the task is to find proper text fragment from the original sentence as the answer of given question based on knowledge understanding.

Transformer neural network is state-of-the-art solution for text analysis and natural language understanding tasks.
With deep multi-layer network structure, transformer neural network is capable of capturing deep semantic information from natural language, which is required by both COVID-19 IR tasks defined in this paper.
Both literature retrieval and question answering tasks are built on the base that the model can well understand the common natural language semantic and COVID-19 domain-specific knowledge inference.
In this paper, we formalize our two tasks into frameworks of two proper natural language questions which can be fitted into transformer neural network.
Based on transformer neural network model,
we implement the aforementioned tasks on a COVID-19 literature dataset, CORD-19.
Using example outputs of our solutions on CORD-19 data, we show the effectiveness of our proposed solutions.
The code for this paper is available at
\
\href{https://github.com/ArusarkaBose/COVID-19}{\texttt{\normalsize{https://github.com/ArusarkaBose/COVID-19}}}

\begin{figure*}[htp]
    \centering
    \includegraphics[width=0.7\linewidth]{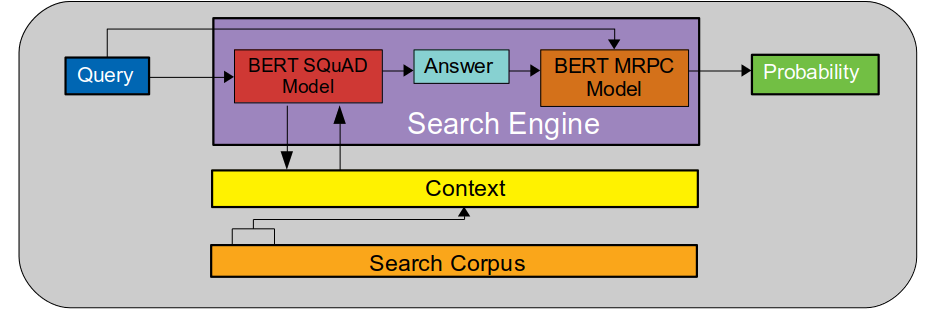}
    \caption{\textbf{COVID-19 Literature Retrieval Model Architecture}}
    \label{fig:Literature Retrieval}
\end{figure*}

\begin{figure*}[htp]
    \centering
    \includegraphics[width=0.7\linewidth]{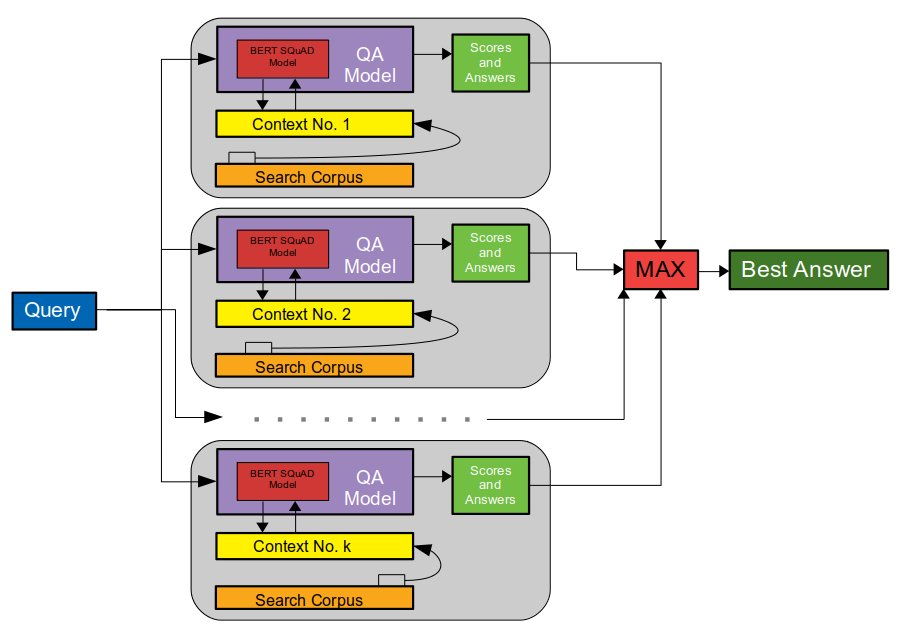}
    \caption{\textbf{COVID-19 Question Answering Model Architecture}}
    \label{fig:my_label}
\end{figure*}

The rest of the paper is organized as follows. Section 2 presents the preliminary of our study. In section 3, we present the state-of-the-art models used in this paper for COVID-19 domain literature information retrieval. We provide the experimental results and discussion in section 4. Finally, section 5 provides the conclusion of this paper. 

\section{\textbf{\Large{Preliminary}}}

Stanford Question Answering Dataset (SQuAD) \cite{Rajpurkar2016SQuAD10} defines a new category of question-answering tasks. From the original textual sentence, a text fragment is picked out as answer of given question. 

The BERT model architecture \cite{devlin2018bert,Vaswani2017AttentionIA} is a multilayer bidirectional Transformer trained on two tasks: 1) predicting randomly masked tokens, and 2) predicting whether two sentences follow each other. 

BERT model can also be fine-tuned with the classifier output layer to do the text sequence classification. And with SQuAD task output layer, the question-answering task can also be achieved by BERT. 

The Hugging Face transformer is built on top of the BERT architecture and provides a unified API for state-of-the-art general-purpose architectures for Natural Language Understanding (NLU) and Natural Language Generation (NLG) with over thousands of pretrained models \cite{Wolf2019HuggingFacesTS} .
  
The COVID-19 Open Research Dataset (CORD-19) \cite{wang2020cord19} is a resource of over 200,000 scholarly articles, including over 90,000 with full text, about COVID-19, SARS-CoV-2, and related coronaviruses. This freely available dataset is provided to the global research community to apply recent advances in NLP and other AI techniques to generate new insights in support of the ongoing fight against this infectious disease.

Although some previous BERT like transformer neural network models, can achieve general text sequence classification and question answering task, the available pre-trained BERT models are trained on common natural language text corpora, and the COVID-19 domain knowledge and semantic is ignored. In this case, we provide COVID-19 customized transformer neural network based IR solutions in this paper to achieve accurate domain IR.   

\section{\textbf{\Large{COVID-19 Domain Literature Information Retrieval}}}    

In this section, we elaborate on the proposed architecture of the Literature-Retrieval and Question-Answering system. As shown in Figure 1, the Literature Retrieval (LR) model is designed to get the top articles in the corpus related to the search query. In Figure 2, the Question-Answering (QA) system provides a correct answer as an output to the search query.  

\textbf{\textsc{\large{Literature Retrieval}}} \ The COVID-19 Literature Retrieval Model comprises a sequential framework of different BERT architectures. The first stage employs the BERT SQuAD model architecture for the selection of documents among the candidate documents in the CORD-19 dataset, with relevance to the given textual query. The BERT MRPC configuration is used in the second stage to rank the selected documents in accordance with similarity to the textual query.

Each candidate document in the CORD-19 dataset is pre-processed into smaller context chunks before passage through the IR system. In each iteration, one of the context chunks is passed along with the textual query through the Literature Retrieval Model in order to generate a probability score which indicates the likelihood of the selected context containing the correct information pertaining to the query.

Figure \ref{fig:Literature Retrieval} illustrates the pathway for a textual query and context pair through the Retrieval Model architecture. The BERT SQuAD neural network model is designed to take as input a query and context pair, and return as output a span of tokens from the context which is the best suitable option for the answer to the input query string.
The first stage of the IR framework employs the BERT SQuAD model to select a word span from the input context which is a possible contender for the answer to the input textual query. The answer thus generated is passed on to the second stage of the framework, along with the originally input textual query.

In the second stage, the BERT MRPC transformer architecture is utilized for the generation of the probability metric. The job of the MRPC architecture is to take in as input a pair of sentences and judge their semantic equivalence on a scale from 0 to 1.
In the proposed Literature Retrieval architecture, the MRPC model serves to adjudicate whether the query, and the selected answer in the first stage involved the same subject matter. The second stage of the IR model outputs the probability score for the chosen context in the current iteration.

Through the described framework, for each candidate document in the CORD-19 dataset, a number of probability scores are obtained corresponding to each context chunk taken out of the document. The maximum of these scores is chosen and assigned as the document-probability-score.
The candidate documents are ranked on the basis of their document-probability-scores and the top ranked serve as the response of the Literature Retrieval system.
%\linebreak \linebreak

\begin{table*}[hbtp!]
    \caption{\textbf{\normalsize{Literature Retrieval System Example Results}}\break (The search corpus taken is a small subset of the CORD-19 dataset and the top 3 results from the IR model has been recorded. The paper ids have been omitted for clarity of reading.)}
    \label{tab:LiteratureRetrievalResults}
    \centering
    \scalebox{0.9}{\begin{tabular}{ |p{0.2cm}|p{19.1cm}|}
        \hline
        \multicolumn{2}{|l|}{\textbf{Query : What are the symptoms?}} \\
        \hline
        1. & Multiple studies have confirmed that BCG is generally safe and can protect children against disseminated disease , including tuberculosis meningitis [ 1 , 2 ] . BCG also provides cross-protection against leprosy [ 3 ] . However , the success of BCG against pulmonary TB in adults is still debated , since randomized clini-cal trials have reported protection efficacy ranging from 0 - 80. \\
        \hline
        2. & However , because reintroduction of ESX-1 into BCG does not restore full virulence , other genetic lesions are also involved [ 56 ] . As such , some strains are more virulent than others in animal models of infection [ 57 ] and also exhibit differential ability to induce adverse reactions ( reactogenicity ) following vaccination in neonates [ 58 ] . \\
        \hline
        3. & ESX-5 is conserved in other pathogenic mycobacteria and reported to facilitate the cell-to-cell spread \\
        \hline
        \multicolumn{2}{|l|}{\textbf{Query : What is the structure?}} \\
        \hline
        1. & Subsequent analyses of multiple vaccine strains have uncovered extensive genome diversity including both deletions and duplications in BCG substrains [ 18 , [ 20 ] [ 21 ] [ 22 ] . The phylogeny established by these molecular methods is consistent with the historical records of BCG dissemination [ 20 , 23 , 24 ] . For example , BCG strains acquired after 1927 exhibit the RD2 deletion , while nRD18 is only deleted in strains obtained after 1933 . Other genomic changes are exclusive to individual daughter strains , and are associated with vaccine production at specific locations [ 22 , 24 ] .A number of molecular techniques have been used to investigate genomic polymorphisms in BCG strains . Early efforts using subtractive hybridization [ 18 ] and spotted oligonucleotide arrays [ 20 , 22 , 25 ] effectively identified large sequence polymorphisms , but lacked the resolution to identify smaller changes . \\
        \hline
        2. & termed B strain , which was responsible for a severe nosocomial outbreak of multidrug resistant TB in humans in Spain [ 50 , 51 ] . However , unlike the three BCG strains , the IS6110 insertion in the M. bovis B strain is located at 75 bp upstream of the start codon of phoP and is in the same orientation as phoP-phoR [ 50 ] . The potential effect of IS6110IS6110 insertion in the phoP promoter in BCG-Russia , -Moreau , and -Japan on phoP expression is described in the ' Discussion ' section . Three other novel phoP-phoR polymorphisms that likely impact their functions were also uncovered by our sequencing analysis . An identical , 11-bp deletion within the ORF of phoR was uncovered in BCG-Sweden and BCG-Birkhaug ( ACCGGACTGGG , nucleotides from 853689 to 853699 , M. tb genome coordinates ) . This deletion changes the amino acid sequence of 54 residues ( residues 432 to 485 ) in the C-terminal of PhoR. This polymorphism is different than the previously described 10 bp deletion within phoR present in BCG-Danish and BCG-Glaxo , which affects residues 91 - 485 [ 24 ] . BCG-Frappier also contains a single nucleotide deletion ( A at 852701 , M. tb genome coordinates ) , causing a frame-shift mutation that affect residues 103 - 485 of PhoR. Together , these results indicate that besides BCG-Danish and BCG-Glaxo , \\
        \hline
        3. & These include genes involved in the biosynthesis of lipid virulence factors PDIMs/PGLs , genes that encode the ESX family type VII secretion system , and the phoP-phoR two-component regulatory system . \\
        \hline
        \multicolumn{2}{|l|}{\textbf{Query : Regions affected?}} \\
        \hline
        1. & The Middle East Respiratory Syndrome ( MERS ) crisis in Korea is coming to an end , but that 's not the end of the story . \\
        \hline
        2. & MERS-CoV is a human betacoronavirus responsible of a severe viral respiratory disease , first identified in Saudi Arabia in 2012 . \\
        \hline
        3. & The WHO experts reminded us that hospital shopping greatly contributed to the rapid spread of MERS in Korea . \\
        \hline
        \multicolumn{2}{|l|}{\textbf{Query : antidote?}} \\
        \hline
        1. & Rigorous evaluation of optimal antimicrobial and other therapeutic strategies to improve the clinical outcomes of this devastating combination is impeded by its sporadic occurrence and fulminant course . Although 838 children with confirmed or suspected 2009 pH1N1 were identified across 35 US pediatric intensive care units ( PICUs ) in 2009 , only 34 cases of MRSA coinfection were reported [ 5 ] .Intravenous vancomycin or clindamycin are recommended as the mainstay of therapy for treatment of hospitalized children with community-acquired ( CA ) pneumonia ( CAP ) if MRSA is suspected [ 6 ] . The addition of a second anti-MRSA agent is controversial [ 7 ] , partly because some combinations of commonly used antibiotics for MRSA , such as linezolid with vancomycin , have shown antagonistic effects in animal models of invasive MRSA infection [ 8 ] and in experimental in vitro assays [ 9 ] . \\
        \hline
        2. & Shock requiring vasopressor support was use of a dopamine infusion \textgreater 5 \textbackslash u00b5g/kg/min or any epinephrine , norepinephrine , or phenylephrine infusion to maintain blood pressure . \\
        \hline
        3. & data from in vitro models also suggest antagonism exists when vancomycin is combined with clindamycin or linezolid [ 7 ] , there is poor agreement between animal and human models of antibiotic antagonism or synergy when treating MRSA [ 38 ] . \\
        \hline
    \end{tabular}}
    \
    \
    \
    \
    \medskip
\end{table*}

\begin{table*}[hbtp!]
\caption{\textbf{\normalsize{Question Answering Results}}\break
    (The search corpus taken is a small subset of the CORD-19 dataset)}
\label{table:Query-Answer}
\centering
\scalebox{0.9}{\begin{tabular}{ |p{4.6cm}|p{14.7cm}|}
\hline
\textbf{\large{Query}} & \textbf{\large{Answer}}\\
\hline
Symptoms of COVID. & Moderate , generalised symmetrical dilation of the ventricular system was present , with complete suppression of contents on FLAIR , indicating that the cerebrospinal fluid was not markedly abnormal . \\
\hline
Which chemicals can be used to kill the virus. & Shock requiring vasopressor support was use of a dopamine infusion \textgreater \ 5 µg/kg/min or any epinephrine , norepinephrine , or phenylephrine infusion to maintain blood pressure . \\
\hline
Coronavirus symptoms. & We excluded patients with preexisting lung disorders ; immune compromise ; mitochondrial , genetic , or neurologic disorders ; and/or preexisting cardiac diseases that increase the risk of infection or respiratory failure [ 24 ] . \\
\hline
What diseases are associated with Coronavirus.    & TEM showed the presence of characteristic small curvilinear lamellar stacks and electron-dense granular material , consistent morphologically with previously described intra-neuronal lipofuscins in cats , making the diagnosis of neuronal ceroid lipofuscinosis highly likely in this case . \\
\hline
Is there any antidote. & Our analysis of the available data on time to presentation and our sensitivity analysis show that although longer time from symptom onset to PICU admission was associated with increased fatality , it was unlikely to explain the association between vancomycin monotherapy and death .\\
\hline
Vaccines for COVID. & Although addition of clindamycin to vancomycin appears prudent in cases with high clinical suspicion of influenza-MRSA coinfection , clindamycin resistance must be monitored .\\
\hline
What is the structure of the virus. & Biochemistry , including fasting ammonia and preprandial bile acids , and haematology were within normal limits . \\
\hline
Is there any medicine for COVID. & Beta-lactam antibiotics were associated with increased S. aureus toxin expression in vitro [ 18 , 19 ] , and vancomycin had negligible effect . \\
\hline
Symptoms. & compare MERS with tuberculosis .\\
\hline
Viral composition. & FK506 binding protein 51 associates with and stabilizes BECN1 by scaffolding the assembly of a hetero-complex involving PHLPP , AKT1 , SKP2 and BECN1 that limits phosphorylation and thus activity of SKP2 .\\
\hline
Where is coronavirus found. & MERS-CoV is a human betacoronavirus responsible of a severe viral respiratory disease , first identified in Saudi Arabia in 2012 .\\
\hline
Effects of COVID-19. & The cat in this case demonstrated generalised and symmetrical brain cortical atrophy and secondary dilation of the ventricular system and intracranial subarachnoid space ( hydrocephalus ex vacuo ) .\\
\hline
Treatment of COVID-19. & Symptomatic management is generally the mainstay of treatment .\\
\hline

\end{tabular}}
\
\
\
\
\medskip

\end{table*}

\textbf{\textsc{\large{Question Answering}}} \ The Question Answering (QA) Model comprises an enhanced version of the first stage of the proposed IR architecture. Similar to the process for the IR framework, the candidate documents from the search corpus (CORD-19) are subdivided into smaller paragraph chunks. Each of the obtained chunks are iteratively streamed through the Question Answering architecture in a two element tuple format along with the textual query string.
Figure \ref{fig:my_label} conceptually describes the complete architecture for the proposed Question Answering Model.

With assistance from the BERT SQuAD framework, the QA model selects a word span from the input context which is the most probable answer for the given textual query, congruent to the process followed in the first stage of the Literature Retrieval Model. The model provides an answer span and a corresponding score indicating the relative likelihood of that answer span. The process is iteratively repeated for the various context chunks of documents present in the search corpus and the answer span with the best overall score is selected as the response of the QA model to the provided textual query.
\linebreak

In order to tackle the non-availability of previously available annotated data to train the transformer architectures for the specific domain in context, we propose semi-manual rule based algorithms to train the neural network models designed for the SQuAD and MRPC tasks.

For training SQuAD architectures, we follow a keyword centric format. We pre-define a list of keywords and a keyword mapping dictionary which maps each of the keywords to some of their synonyms. For example if the list of keywords is chosen as,
\[[`structure',`symptoms']\]
then one of the possible representations of the mapping dictionary can be,
\[\{`structure': [`structure',`constituents',`composition'],\]
\[`symptoms': [`symptoms',`effects',`diseases']\}\]
For selecting the questions in the required \textit{(question, context)} pair for the SQuAD datafile format, we pre-define a list of queries.

Similar to the preprocessing step for our proposed model architectures, before proceeding with the creation of training datasets, the available documents to be used for training are divided into chunks of contexts. On selecting a context from the available chunks, for each of the keywords in our defined list of keywords, we choose the lines from that context which pertains to the same subject-matter as the keyword does. Unlike all the other steps in our training pipeline, this particular step is a manual one and serves as the main temporal bottleneck of the proposed pipeline. 

In order to choose the relevant lines, we take the assistance of the scispaCy named entity recognition models \cite{neumann-etal-2019-scispacy}. For each \textit{(context, keyword)} pairs, we
\begin{itemize}
    \item Find all the named entities in the context using one of the scispaCy NER models.
    \item Choose the entities having relevance to the current keyword.
    \item Access the subtrees of the chosen entities and select the lines in the context that contains those entities.
\end{itemize}
In this way, a dictionary consisting of the contexts and the lines selected from them for each keyword in the keyword list, can be formed.

For any of the queries in our pre-defined list of queries, the query is scanned for the presence of any of the words contained in the keyword mapping dictionary and, if found, that word is mapped to the appropriate keyword in the keyword list. Thus, a list of keywords that are present in the sentence can be collected. The SQuAD training datafile consists of \textit{(question, answer, context)} triplets. In our proposed pipeline, these creation of these triplets is implemented by the following steps:
\begin{itemize}
    \item Choose a query from the list of queries.
    \item Choose a context chunk from the available chunks.
    \item Extract the keywords present in the query and choose the appropriate lines from the context as the answer to the query with referral from the dictionary previously created.
\end{itemize}
%\
%\linebreak

The Microsoft Research Paraphrase Corpus (MRPC) task consists of checking whether two sentences are semantically equivalent \cite{Dolan2005AutomaticallyCA}. The MRPC trainfile expects sentence sample pairs with scores of 1 or 0 indicating semantic similarity and dissimilarity respectively. In order to create our question-answer sentence pairs, we salvage the previously created SQuAD trainfile. For creating the positive samples (the samples with a score of 1),
\begin{itemize}
    \item Select any query and one of its corresponding answers from the SQuAD trainfile
    \item Split the selected answer into its constituent sentences. Our proposed framework uses the NLTK punkt tokenizer for this purpose.
    \item Choose any of the constituent sentences and the selected query as a positive sample pair
\end{itemize}

The steps followed for the creation of the negative samples are sightly different. We start with selecting a query and a context from the SQuAD trainfile and subsequently select the lines from that context that are not present among the answers for the query. As mentioned in the preceding paragraph, the punkt tokenizer is used for sentence extraction. The generation of the positive and negative samples is followed by a shuffling of the samples to prevent any bias in training.

\section{\textbf{\Large{Experiment Results}}}

Table \ref{tab:LiteratureRetrievalResults} and \ref{table:Query-Answer} summarize our experiments for the two proposed architectures. In order to exhibit the efficiency of the presented models, we select a very small subset of the CORD-19 dataset as the search corpus for the experiments.

In table \ref{tab:LiteratureRetrievalResults}, we choose four example user queries for the COVID-19 Literature Retrieval solution. The framework returns a list of publication IDs and the selected excerpt for each publication. The list is arranged in decreasing order of relevance to the textual query provided as input. We select the top 3 results and present them in the table for each of the four example queries. The publication IDs have been omitted for cogency. The proposed solution provided the results which have high semantic and knowledge relevance with given queries.

In table \ref{table:Query-Answer}, we list some user queries, by using them as questions, our COVID-19 question answering solution is able to select proper text fragments from the search corpus as answers. The predicted answers are sufficiently relevant to the input questions and contain the effective knowledge or information which is important in order to answer the given textual queries.

Though the number of queries our training samples contain is limited, unseen queries can be answered by using word embeddings such as Word2Vec \cite{mikolov2013efficient} or GloVe \cite{Pennington2014GloveGV} to calculate the similarity between the words in the unseen query and the keywords in our keyword list. If the similarity with any of the keywords is greater than a specified cutoff, that particular word in unseen query can be replaced with that keyword and can subsequently be processed with an accuracy closer to that of the queries available in the training scripts.
%\linebreak %\linebreak
For unseen questions with proper nouns like the names of countries, chemicals, etc., for example, 
\[``Coronavirus\ cases\ in\ Australia"\]
or, 
\[``ChAdOx1\ nCoV-19\ as\ COVID\ vaccine"\] 
exact match of the proper noun with any of the words in the context, can be used as a measure of score to rank the papers, along with the Literature Retrieval architecture scores. Similar is the workaround for the proposed Question-Answering model.
%\linebreak %\linebreak

\section{\textbf{\Large{Conclusion}}}
In this paper,
we define a novel COVID-19 Domain Literature Information Retrieval framework which includes two tasks, COVID-19 literature retrieval, and question answering.
Based on Transformer Neural Network model, we propose solutions for these two tasks respectively.
In our solutions, the COVID-19 domain knowledge and particularity are considered in IR process.
In this case, our solutions achieve good IR results on CORD-19 dataset.

\printbibliography

@article{wang2020cord19,
    title={CORD-19: The COVID-19 Open Research Dataset},
    author="Lucy Lu Wang and Kyle Lo and Yoganand Chandrasekhar and Russell Reas and Jiangjiang Yang and Doug Burdick and Darrin Eide and Kathryn Funk and Yannis Katsis and Rodney Kinney and Yunyao Li and Ziyang Liu and William Merrill and Paul Mooney and Dewey Murdick and Devvret Rishi and Jerry Sheehan and Zhihong Shen and Brandon Stilson and Alex Wade and Kuansan Wang and Nancy Xin Ru Wang and Chris Wilhelm and Boya Xie and Douglas Raymond and Daniel S. Weld and Oren Etzioni and Sebastian Kohlmeier",
    year={2020},
    eprint={2004.10706},
    archivePrefix={arXiv},
    primaryClass={cs.DL}
}

@article{devlin2018bert,
  title={BERT: Pre-training of Deep Bidirectional Transformers for Language Understanding},
  author={Devlin, Jacob and Chang, Ming-Wei and Lee, Kenton and Toutanova, Kristina},
  journal={arXiv preprint arXiv:1810.04805},
  year={2018}
}

@article{Vaswani2017AttentionIA, 
title={Attention is All you Need}, 
author={Ashish Vaswani and Noam Shazeer and Niki Parmar and Jakob Uszkoreit and Llion Jones and Aidan N. Gomez and Lukasz Kaiser and Illia Polosukhin}, 
journal={ArXiv}, 
year={2017}, 
volume={abs/1706.03762} 
}

@article{Wolf2019HuggingFacesTS,
  title={HuggingFace's Transformers: State-of-the-art Natural Language Processing},
  author={Thomas Wolf and Lysandre Debut and Victor Sanh and Julien Chaumond and Clement Delangue and Anthony Moi and Pierric Cistac and Tim Rault and R'emi Louf and Morgan Funtowicz and Jamie Brew},
  journal={ArXiv},
  year={2019},
  volume={abs/1910.03771}
}

@inproceedings{neumann-etal-2019-scispacy,
    title = "{S}cispa{C}y: {F}ast and {R}obust {M}odels for {B}iomedical {N}atural {L}anguage {P}rocessing",
    author = "Neumann, Mark  and
      King, Daniel  and
      Beltagy, Iz  and
      Ammar, Waleed",
    booktitle = "Proceedings of the 18th BioNLP Workshop and Shared Task",
    month = aug,
    year = "2019",
    address = "Florence, Italy",
    publisher = "Association for Computational Linguistics",
    url = "https://www.aclweb.org/anthology/W19-5034",
    doi = "10.18653/v1/W19-5034",
    pages = "319--327",
    eprint = {arXiv:1902.07669},
}

@article{Rajpurkar2016SQuAD10, 
title={SQuAD: 100, 000+ Questions for Machine Comprehension of Text}, 
author={Pranav Rajpurkar and Jian Zhang and Konstantin Lopyrev and Percy Liang}, 
journal={ArXiv}, 
year={2016}, 
volume={abs/1606.05250} 
}

@inproceedings{Dolan2005AutomaticallyCA, 
title={Automatically Constructing a Corpus of Sentential Paraphrases}, 
author={William B. Dolan and Chris Brockett}, 
booktitle={IWP@IJCNLP}, 
year={2005} 
}

@misc{mikolov2013efficient,
    title={Efficient Estimation of Word Representations in Vector Space},
    author={Tomas Mikolov and Kai Chen and Greg Corrado and Jeffrey Dean},
    year={2013},
    eprint={1301.3781},
    archivePrefix={arXiv},
    primaryClass={cs.CL}
}

@inproceedings{Pennington2014GloveGV, 
title={Glove: Global Vectors for Word Representation}, 
author={Jeffrey Pennington and Richard Socher and Christopher D. Manning},
booktitle={EMNLP}, 
year={2014} 
}
\end{document}